\documentclass[letter,nofootinbib,12pt,superscriptaddress]{revtex4-2}

\usepackage[colorlinks,citecolor=blue]{hyperref}
\usepackage{amsmath}
\usepackage{mathrsfs}
\usepackage{bbm}
\usepackage{amsfonts}
\usepackage{amssymb}
\usepackage{latexsym}
\usepackage{graphicx}
\usepackage[english]{babel}
\usepackage{multirow}
\usepackage{float}
\usepackage{url}
\usepackage{slashed}
\usepackage[compat=1.1.0]{tikz-feynman}
\usepackage{orcidlink}

\def\be{\begin{equation}}
\def\ee{\end{equation}}
\def\bea{\begin{eqnarray}}
\def\eea{\end{eqnarray}}

\def\mF{\mathcal{F}}

\def\zd{\mathrm{d}}

\begin{document}

\title{Thermodynamic of the $f(Q)$ universe}

\author{Haomin Rao}
\email{rhm137@mail.ustc.edu.cn}
\affiliation{School of Fundamental Physics and Mathematical Sciences,
Hangzhou Institute for Advanced Study, UCAS, Hangzhou 310024, China}
\affiliation{University of Chinese Academy of Sciences, 100190 Beijing, China}

\author{Chunhui Liu}
\email{liuchunhui22@mails.ucas.ac.cn}
\affiliation{School of Fundamental Physics and Mathematical Sciences,
Hangzhou Institute for Advanced Study, UCAS, Hangzhou 310024, China}
\affiliation{University of Chinese Academy of Sciences, 100190 Beijing, China}
\affiliation{Institute of Theoretical Physics, Chinese Academy of Sciences, Beijing 100190, China}

\author{Chao-Qiang Geng}
\email{cqgeng@ucas.ac.cn}
\affiliation{School of Fundamental Physics and Mathematical Sciences,
Hangzhou Institute for Advanced Study, UCAS, Hangzhou 310024, China}
\affiliation{Department of Physics and Synergetic Innovation Center for Quantum Effects and Applications, Hunan Normal University, Changsha, Hunan 410081, China}

\begin{abstract}
{\color{white} nothing}

We investigate thermodynamics of apparent horizon in the $f(Q)$ universe with trivial and nontrivial connections.
We first explore the perspectives of the first law, generalized second law and $P-V$ phase transition with trivial connection.
We show that the lowest-order correction of entropy has the same form as that in loop quantum gravity,
and the critical exponents of the phase transition caused by the lowest-order correction are consistent with those in mean field theory.
We then examine the thermodynamic implication of nontrivial connections.
We find that nontrivial connections in the $f(Q)$ universe imply non-equilibrium states from the perspective of thermodynamics.
\end{abstract}

\maketitle

\newpage

\section{Introduction}\label{Introduction}

The four laws of black hole mechanics in general relativity (GR) are very similar to the four laws of thermodynamics \cite{Bardeen:1973gs}.
Classically, black holes seem to have neither temperature nor entropy.
However, the area law of black hole entropy can be derived from information theory \cite{Bekenstein:1973ur}
and the black hole temperature can be derived from quantum field theory in curved spacetime \cite{Hawking:1975vcx}.
This suggests that the black hole thermodynamics is not just an analogy,
but reveals a hidden relationship between gravity, thermodynamics and quantum theory \cite{Jacobson:1995ab,Jacobson:2015hqa,Padmanabhan:2009vy,Wald:1999vt}.
Following this idea, more thermodynamic concepts such as phase transition have been introduced into the study of black holes
\cite{Kubiznak:2016qmn,Hawking:1982dh,Altamirano:2014tva}.
In addition, it has been found that black holes in different gravity models have different thermodynamic behaviors \cite{Cai:2001dz,Cai:2003kt,Akbar:2006mq,Miao:2011ki},
indicating that black hole thermodynamics contains information about gravity models.

The universe is another neat physical system closely related to gravity.
Inspired by the black hole thermodynamics,
people found that the evolution equations of the universe can be written in the form of thermodynamic laws \cite{Padmanabhan:2003gd,Cai:2005ra}.
Since the temperature of the horizon in the universe can also be derived by quantum methods \cite{Gibbons:1977mu,Cai:2008gw,DiCriscienzo:2009kun,Zhu:2009wa},
the thermodynamics of the universe once again reveals the hidden relationship between gravity and quantum theory.
Even more surprising is that in some gravity models the horizon entropy of the black hole
and the horizon entropy of the universe have exactly the same form \cite{Akbar:2006kj,Cai:2006rs,Kong:2021dqd}.
This reflects that the thermodynamics of the universe and the thermodynamics of black holes are not independent in a given gravity model.
Clearly, it is necessary and attractive to discuss the thermodynamics of the universe in various gravity models \cite{Akbar:2006er,Gong:2007md,Bamba:2009id,Bamba:2011pz,Bamba:2009ay,Bamba:2009gq,Bamba:2011jq,Bamba:2010kf}.

Symmetric teleparallel gravity (STG) is a newly popular modified gravity scheme that identifies gravity as non-metricity rather than curvature or torsion.
The simplest STG model classically is  equivalent to GR \cite{Nester:1998mp,BeltranJimenez:2019esp,Capozziello:2022zzh},
which provides us with another way to modify GR, that is, to modify the simplest STG model.
The most eye-catching and interesting modified STG model is the $f(Q)$ model \cite{BeltranJimenez:2017tkd,Heisenberg:2023lru},
and its cosmological applications have been extensively studied in the literature \cite{BeltranJimenez:2019tme,Atayde:2021pgb,Oliveros:2023mwl,Khyllep:2022spx,Zhao:2021zab,Hu:2023ndc}.
More recent studies have shown that in the $f(Q)$ model,
there are three different connections that satisfy cosmological symmetry in the flat universe \cite{Hohmann:2021ast,Heisenberg:2022mbo,Hohmann:2020zre,Gomes:2023hyk}.
And different connections can result in different background evolutions as well as different perturbation behaviors
\cite{Dimakis:2022rkd,Shi:2023kvu,Gomes:2023tur,Rao:2023nip}.
This makes the flat universe in the $f(Q)$ model more complex but with richer phenomenology.

So far, the thermodynamics of the $f(Q)$ universe and the thermodynamic implication of different connections remain to be explored.
In this paper, we will investigate the thermodynamics of $f(Q)$ universes with different connections.
Since the case of trivial connection has attracted overwhelming attention in past studies of the $f(Q)$ universe,
the thermodynamics of the $f(Q)$ universe with trivial connection is also the primary target in this paper.
We will explore it from the perspectives of the first law, generalized second law and $P-V$ phase transition.
After that, we will examine the thermodynamic implication of nontrivial connections.
This will provides the first glimpse into how different connections affect the thermodynamics of the universe in the same gravity model.

The present paper is organized as follows.
In section \ref{SecfQ}, we briefly introduce the $f(Q)$ gravity and its flat universe background with trivial and nontrivial connections.
In section~\ref{Sectrivial}, we analyse the thermodynamics of the flat universe with trivial connection from the following three aspects.
In subsection \ref{Secfirstlaw}, we examine the first law of thermodynamics and derive the area law of entropy for the most general $f(Q)$ model.
Subsequently, we briefly discuss the restrictions on the $f(Q)$ universe brought by the generalized second law in subsection \ref{Secsecondlaw}.
In subsection \ref{Sectransition}, we study the $P-V$ phase transition in a simple $f(Q)$ model and calculate all critical exponents.
Finally, in section \ref{Secnontrivial}, we show the thermodynamic implication of nontrivial connections from the perspective of the first law.

In this paper, we adopt the unit $ G=1$ and the signature $(-,+,+,+)$.
The spacetime indices are denoted by Greek indices $\mu, \nu, \rho,...=0, 1, 2, 3$ and the spatial indices are represented by Latin indices $i, j, k,...=1, 2, 3$.
In addition, we distinguish the spacetime affine connection $\Gamma^{\rho}_{~\mu\nu}$
and its associated covariant derivative $\nabla$ from the Levi-Civita connection $\mathring{\Gamma}^{\rho}_{~\mu\nu}$
and its associated covariant derivative $\mathring{\nabla}$, respectively.

\section{$f(Q)$ gravity and its cosmology}\label{SecfQ}

The so-called STG theory is formulated in a spacetime endowed with a metric $g_{\mu\nu}$ and an affine connection $\Gamma^{\rho}_{~\mu\nu}$,
which is curvature-free and torsion-free. Without curvature and torsion, the gravitation is identified with non-metircity
\be\label{nonmetricity}
Q_{\rho\mu\nu}=\nabla_{\rho}g_{\mu\nu}.
\ee
The connection $\Gamma^{\rho}_{~\mu\nu}$ on such a geometry can be  generally expressed as
\be\label{STGconnection}
\Gamma^{\rho}_{~\mu\nu}=\frac{\partial x^{\rho}}{\partial y^{\sigma}}\partial_{\mu}\partial_{\nu}y^{\sigma}.
\ee
As a result, we can regard $g_{\mu\nu}$ and $ y^{\mu} $ as the basic variables of the STG theory.

The simplest STG model is the so-called symmetric teleparallel equivalent of general relativity (STEGR) model, whose action is
\be\label{STEGR}
I_{\text{STEGR}}=\frac{1}{16\pi}\int \zd^4x \sqrt{-g}\, Q+I_{m},
\ee
with the non-metricity scalar $Q$ defined as
\be\label{nonmetricity}
Q=P^{\rho\mu\nu}Q_{\rho\mu\nu}
=-\frac{1}{4}Q_{\rho\mu\nu}Q^{\rho\mu\nu}+\frac{1}{2}Q_{\rho\mu\nu}Q^{\mu\nu\rho}+\frac{1}{4}Q_{\mu}Q^{\mu}-\frac{1}{2}Q_{\mu}\tilde{Q}^{\mu},
\ee
where the non-metricity conjugate is $P^{\rho\mu\nu}=-\frac{1}{4}Q^{\rho\mu\nu}+\frac{1}{2}Q^{(\mu\nu)\rho}+\frac{1}{4}(Q^{\rho}-\tilde{Q}^{\rho})g^{\mu\nu}-\frac{1}{4}g^{\rho(\mu}Q^{\nu)}$ with
$Q_{\mu}=Q_{\mu\nu}{}^{\nu}$, $\tilde{Q}_{\mu}=Q^{\nu}{}_{\nu\mu}$.
Since we have the identity
$\mathring{R}\equiv Q-\mathring{\nabla}_{\mu}(Q^{\mu}-\tilde{Q}^{\mu})$,
the action~(\ref{STEGR}) is identical to the Einstein-Hilbert action up to a surface term,
where the curvature scalar $\mathring{R}$ is defined by the Levi-Civita connection.
Since the surface term in the action does not affect the equations of motion,
we say that  STEGR is equivalent to GR at the level of equations of motion.

This equivalence provides us with another way to modify GR, which is to modify the STEGR model within the STG framework.
Along this line, a variety of modified STG models have been proposed.
The most interesting one is the $f(Q)$ model,
which generalizes $Q$ in the action (\ref{STEGR}) to a smooth function $f(Q)$, given by
\be\label{model}
I=\frac{1}{16\pi}\int \zd^4x \sqrt{-g}\,f(Q)+I_m,
\ee
The equations of motion follow from the variations with respect to $g_{\mu\nu}$ and $y^{\mu}$ are
\bea
\label{Eom1}
f_{Q}\mathring{G}^{\mu\nu}+\frac{1}{2}(Qf_{Q}-f)g^{\mu\nu}
+2\nabla_{\rho}f_{Q}P^{\rho\mu\nu}&=&8\pi T^{\mu\nu},
\\ \label{Eom2}
\nabla_{\mu}\nabla_{\nu}\left(\sqrt{-g}f_{Q}P^{\mu\nu}{}_{\rho}\right)&=&0,
\eea
where
$f_{Q}=\frac{\zd f(Q)}{\zd Q}$, $\mathring{G}^{\mu\nu}=\mathring{R}^{\mu\nu}-\frac{1}{2}\mathring{R}g^{\mu\nu}$ is the Einstein tensor,
$T^{\mu\nu}=\frac{2}{\sqrt{-g}}\frac{\delta I_m}{\delta g_{\mu\nu}}$ is the energy-momentum tensor of matter.
It can be verify that if $f(Q)=Q$, Eq.~(\ref{Eom1}) becomes the Einstein field equation, while Eq.~(\ref{Eom2}) is reduced to an identity.

After clarifying the $f(Q)$ model, let's briefly introduce its cosmology.
In the flat universe, the metric can be expressed in rectangular coordinate as
\be\label{FRWmetric}
\zd s^{2}=-\zd t^2+a^2(t)\delta_{ij}\zd x^{i}\zd x^j,
\ee
where $a=a(t)$ is the scale factor.
Eq.~(\ref{FRWmetric}) comes from the symmetry requirement that the metric is homogeneous and isotropic, i.e. $\mathcal{L}_{\xi}g_{\mu\nu}=0$,
where $\mathcal{L}_{\xi}$ is the Lie derivative and $\xi$ represents all six Killing vector fields in the flat universe.
In the STG theory, the connection cannot be completely determined by the metric.
As suggested in Refs.~\cite{Hohmann:2021ast,Heisenberg:2022mbo,Hohmann:2020zre,Gomes:2023hyk}, it is natural to further require that the connection is also homogeneous and isotropic, that is,
\be\label{FRWconnection0}
\mathcal{L}_{\xi}{\Gamma}^{\rho}_{~\mu\nu}={\nabla}_{\mu}{\nabla}_{\nu}\,\xi^{\rho}=0.
\ee
It can be obtained from Eq.~(\ref{FRWconnection0}) that
the non-zero components of the connection ${\Gamma}^{\rho}_{~\mu\nu}$ are
\be\label{FRWconnection}
\Gamma^{0}_{~00}=K_{1}~,~\Gamma^{0}_{~ij}=a^{2}K_{2}\,\delta_{ij}~,~\Gamma^{i}_{~0j}=\Gamma^{i}_{~j0}=K_{3}\,\delta_{ij},
\ee
with $\{K_1, K_2, K_3\}$ having three branch solutions, given by
\bea
& &
\text{branch 1 :}~ K_1=\gamma~,~K_2=0~,~K_3=0,\label{branch1}
\\ & &
\text{branch 2 :}~ K_1=\dot{\gamma}/\gamma+\gamma~,~K_2=0~,~K_3=\gamma, \label{branch2}
\\ & &
\text{branch 3 :}~ K_1=-(\dot{\gamma}/\gamma+2H),~K_2=\gamma~,~K_3=0,\label{branch3}
\eea
where $H=\dot{a}/a$ is the Hubble rate, $\gamma=\gamma(t)$ is a function of  time $t$
and the superscript ``dot" represents the derivative with respect to $t$.
It can be seen that even if the metric and connection have been required to be homogeneous and isotropic,
the form of the cosmological background is not unique.
This is a unique feature of the STG universe, which does not appear in $f(R)$ and $f(T)$.

Putting the metric (\ref{FRWmetric}) and the connection (\ref{FRWconnection}) into Eqs.~(\ref{Eom1}) and (\ref{Eom2}),
we obtain the background equations as
\bea
& &\label{beom1}
3f_{Q}H^{2}+\frac{1}{2}(f-Qf_{Q})+\frac{3}{2}(K_{3}-K_{2})\dot{Q}f_{QQ}=8\pi\rho_{m},
\\ & &\label{beom2}
f_{Q}(2\dot{H}+3H^{2})+\frac{1}{2}(f-Q f_{Q})+\frac{1}{2}(4H-K_{2}-3K_{3})\dot{Q}f_{QQ}=-8\pi p_{m},
\\ & &\label{beom3}
(K_{2}-K_{3})\ddot{f}_{Q}+\left(HK_{2}-3HK_{3}-2K_{1}K_{2}\right)\dot{f}_{Q}=0,
\eea
where  $Q=-6H^{2}$ in branch 1 and $Q=-6H^{2}+9H\gamma+3\dot{\gamma}$ in branch 2 and branch 3, $f_{QQ}=\zd f_{Q}/\zd Q$,
$\rho_{m}$ and $p_{m}$ are the energy density and pressure of matter, respectively.
Eqs.~(\ref{beom1})-(\ref{beom3}) show that the connection can affect the evolution of the cosmological background.
In fact, this feature has only recently been taken seriously.
Almost all early studies of the $f(Q)$ universe have assumed the trivial connection $\Gamma^{\rho}_{~\mu\nu}=0$.
This is equivalent to considering only the branch 1 solution with $\gamma=0$, which is only a special solution of the condition (\ref{FRWconnection0}).
In this case, Eqs.~(\ref{beom1}) and (\ref{beom2}) can be simplified to
\bea
& & \label{beom01} 3f_{Q}H^{2}+\frac{1}{2}(f-Qf_{Q})=8\pi\rho_{m},\\
& & \label{beom02} f_{Q}(2\dot{H}+3H^{2})+\frac{1}{2}(f-Q f_{Q})+2H\dot{f}_{Q}=-8\pi p_{m},
\eea
while Eq.~(\ref{beom3}) degenerates into an identity.
Eq.~(\ref{beom01}) and (\ref{beom02}) are exactly the background equations adopted in most studies of the $f(Q)$ universe.

\section{Thermodynamic with trivial connection}\label{Sectrivial}
After a brief introduction to $f(Q)$ gravity and its flat universe, in this section, we study the thermodynamics of the $f(Q)$ universe.
Since the case of trivial connection $\Gamma^{\rho}_{~\mu\nu}=0$ has attracted overwhelming attention in the studies of the $f(Q)$ universe,
we only focus on the thermodynamics of the universe with trivial connection in this section.
We will explore the thermodynamics from the perspectives of the first law, generalized second law and $P-V$ phase transition.
The thermodynamic effects of nontrivial connections are left to the next section.

\subsection{First Law of Thermodynamics}\label{Secfirstlaw}
Similar to the black hole horizon, there is also thermodynamics at the cosmological horizon.
For the convenience of discussions, we express the metric (\ref{FRWmetric}) as the following form with obvious spherical symmetry
\be\label{sphericalmetric}
\zd s^{2}=h_{\alpha\beta}\zd z^{\alpha}\zd z^{\beta}+R^{2}\zd\Omega_{2}^{2},
\ee
where $R=a|\vec{x}|$ is physics radius, $\zd\Omega_{2}^{2}$ is the line element of a 2-dimensional sphere with unit radius,
$z^{\alpha}=(t,|\vec{x}|)$,  and $h_{\alpha\beta}=\text{diag}\{-1,a^2\}$ is the induced metric on the $z$-plane.
Since the event horizon depends on the whole history of the universe,
the apparent horizon is the more natural horizon, which is a trapped surface with a vanishing expansion of $h^{\alpha\beta}{\partial_\alpha}R{\partial_\beta}R=0$.
The apparent horizon of the flat universe can be easily found as
\be\label{horizonR}
R_{A}=H^{-1}.
\ee
which is exactly equal to the Hubble horizon.
Then, the surface gravity on the apparent horizon (\ref{horizonR}) can be obtained as
\be\label{surfacegravity}
\kappa=\frac{1}{2\sqrt{-h}}\partial_{\alpha}\left(\sqrt{-h}h^{\alpha\beta}\partial_{\beta}R\right)_{R_{A}}=-\frac{1}{R_A}(1-\frac{\dot{R}_A}{2})
\ee

Inspired by the thermodynamics of black holes, the temperature of the apparent horizon in the flat universe
is defined by the surface gravity $\kappa$ as
\be\label{temperature}
T=\frac{|\kappa|}{2\pi}=\frac{1}{2\pi R_{A}}\left(1-\frac{\dot{R}_{A}}{2}\right).
\ee
The temperature (\ref{temperature}) is consistent with the one obtained
by the Hamilton-Jacobi tunneling method for the apparent horizon \cite{DiCriscienzo:2009kun}.
Following Refs.~\cite{Bak:1999hd,Hayward:1998ee,Hayward:1997jp}, we define the work density by $W=-\frac{1}{2}h_{\alpha\beta}T^{\alpha\beta}$.
Using background equations (\ref{beom01}) and (\ref{beom02}) we can find the work density in the $f(Q)$ flat universe as
\be\label{workdensity}
W=\frac{1}{8\pi}\Big[f_Q(\dot{H}+3H^2)+\frac{1}{2}(f-Qf_Q)+H\dot{f}_{Q}\Big].
\ee
In addition, the total energy of matter inside the apparent horizon can be easily obtained from the background equation (\ref{beom01})
\be\label{energy}
E=\rho_{m} V=\frac{R_Af_Q}{2}+\frac{R_A^3}{12}(f-Qf_Q),
\ee
where $V=\frac{4}{3}\pi R_{A}^{3}$ is the thermodynamic volume, which is also the physical volume within the apparent horizon in the flat universe.
Finally, from the internal energy $U=-E$ and the thermodynamic pressure $P=W$, the standard first law of thermodynamics can be established
\be\label{firstlaw}
\zd U=T\zd S-P\zd V,
\ee
where the entropy $S$ can be integrated as a function of the horizon area
\be\label{entropy}
S(A)=\frac{1}{4}\int\left( f_{Q}+2Q f_{QQ}\right)\zd A,
\ee
with $Q=-24\pi A^{-1}$, where $A=4\pi R_{A}^{2}$ is the thermodynamic area as well as the physical area of the apparent horizon.
This means that the thermodynamics of the apparent horizon is always the equilibrium thermodynamics  in the $f(Q)$ universe with trivial connection.

In order to explore the horizon entropy (\ref{entropy}) more intuitively,
let’s take a look at the entropy (\ref{entropy}) in some simple $f(Q)$ models.
The simplest $f(Q)$ model is the STEGR model, that is, $f(Q)=Q$.
The entropy of the STEGR model can be obtained by Eq.~(\ref{entropy}) as
\be\label{entropy0}
S(A)=\frac{A}{4},
\ee
which is exactly the Bekenstein-Hawking entropy.
Next, from the perspective of the Taylor expansion,
the $f(Q)$ model retaining the lowest-order correction is $f(Q)=Q+\lambda\, Q^{2}$.
The entropy in this model can be obtained by Eq.~(\ref{entropy}) as
\be\label{entropy1}
S(A)=\frac{A}{4}-36\pi\lambda{\rm ln}\frac{A}{A_0},
\ee
where $A_{0}$ is an integration constant.
Compared with the Bekenstein-Hawking entropy, the entropy (\ref{entropy1}) has an additional logarithmic correction.
Such a logarithmic correction can also arise from
the lowest-order correction of loop quantum gravity due to thermal equilibrium fluctuations and quantum fluctuations
\cite{Meissner:2004ju,Ghosh:2004rq,Chatterjee:2003uv,Banerjee:2009tz,Modak:2008tg,MohseniSadjadi:2010nu}.
Next, if we continue to follow the Taylor expansion and keep it to the second-order,
then the function $f(Q)$ can be expressed as $f(Q)=Q+\lambda\, Q^{2}+c\, Q^{3}$.
The entropy in this model can be obtained by Eq.~(\ref{entropy}) as
\be\label{entropy2}
S(A)=\frac{A}{4}-36\pi\lambda\ln\frac{A}{A_0}-\frac{2160\pi^{2}c}{A}.
\ee
Comparing with Eq.~(\ref{entropy1}), it is found that the second-order correction of entropy is a power law correction.
The power law correction of entropy also appears in the second-order correction of loop quantum gravity
\cite{Banerjee:2009tz,Modak:2008tg,MohseniSadjadi:2010nu}
as well as the entanglement of quantum fields between in and out the horizon \cite{Das:2007mj,Radicella:2010ss,Sheykhi:2011egx}.
A succession of coincidences seems to suggest that the $f(Q)$ model is linked to the effects of quantum gravity.
However, this topic is beyond the scope of this paper and is left for future research.

\subsection{Generalized Second Law of Thermodynamics}\label{Secsecondlaw}

In the previous subsection, we derive the horizon entropy from the first law of thermodynamics.
It follows that one might wonder what would happen if we required that the total entropy does not decrease.
In this subsection, we investigate the generalized second law of thermodynamics in the $f(Q)$ universe,
which asserts that the entropy of the horizon plus the entropy of matter within the horizon does not decrease.

Suppose that there are $i=1,2,...,n$ components in the universe, and the $i$-th component follows continuity equation
\be
\rho_{i}+3H(\rho_{i}+p_{i})=q_{i},
\ee
where $\rho_{i}$ and $p_{i}$ are the energy density and pressure of the $i$-th component, respectively,
and  $q_i$ is the interaction term, which reflects that energy can be exchanged between different components.
The total energy density $\rho_{m}=\sum\rho_{i}$ and pressure $p_{m}=\sum p_{i}$ must satisfy the continuity equation, which means
\be\label{continuityEq}
\dot{\rho}_{m}+3H(\rho_{m}+p_{m})=0~~~\Rightarrow~~~\sum_{i=1}^{n}q_{i}=0.
\ee
Based on the discussion of Refs.~\cite{Davies:1987ti,Clifton:2007tn,MohseniSadjadi:2005ps},
we apply the first law of thermodynamics to each component
\be\label{firstlawmatter}
\zd E_{i}=T_{i}\zd S_{i}-p_{i}\zd V,
\ee
where $E_{i}=\rho_{i}V$ is the total energy of the $i$-th component within the horizon, $T_i$ is the temperature of the $i$-th component,
and $S_i$ is the total entropy of the $i$-th component within the horizon.
From the first law (\ref{firstlawmatter}), we can derive the change rate of the total entropy of matter
\be
\dot{S}_{m}=\frac{4\pi}{3H^{3}}\sum_{i=1}^{n}\frac{q_{i}}{T_{i}}-4\pi\frac{\dot{H}+H^{2}}{H^{4}}\sum_{i=1}^{n}\frac{\rho_{i}+p_{i}}{T_{i}},
\ee
where $S_{m}=\sum S_{i}$ is the total entropy of matter within the horizon.
On the other hand, the change rate of the horizon entropy can be obtained from Eq.~(\ref{entropy}), given by
\be
\dot{S}(A)=-\frac{2\pi\dot{H}}{H^{3}}(f_{Q}+2Qf_{QQ}).
\ee
Finally, the change rate of the total entropy within the horizon can be found
\be\label{entropychange}
\dot{S}_{t}=\frac{4\pi}{3H^{3}}\sum_{i=1}^{n}\frac{q_{i}}{T_{i}}-4\pi\frac{\dot{H}+H^{2}}{H^{4}}\sum_{i=1}^{n}\frac{\rho_{i}+p_{i}}{T_{i}}
-\frac{2\pi\dot{H}}{H^{3}}(f_{Q}+2Qf_{QQ}),
\ee
where $S_{t}=S_{m}+S(A)$ is the total entropy within the horizon.
The so-called generalized second law of thermodynamics requires $\dot{S}_{t}\geq0$, which is what we are going to discuss.

The composition of the real universe is very complex, which inevitably makes the analysis of the second law $\dot{S}_{t}\geq0$ complicated and bloated.
In order to make the analysis of the second law more concise and intuitive, we study the simplified toy universe in this subsection.
Firstly, we assume that all components have the same temperature, that is, $T_{i} = T_m$ for any $i=1,2,...,n$.
Then, by combining the continuity equation (\ref{continuityEq}) and the background equations (\ref{beom01}) and (\ref{beom02}),
the generalized second law can be simplified to
\be
\dot{S}_{t}=\frac{\dot{H}}{H^{4}T_{m}}(\dot{H}+H^{2}-2\pi H T_{m})(f_{Q}+2Qf_{QQ})\geq0.
\ee
Secondly, we also assume that the temperature of matter is equal to the temperature of the horizon.
This assumption has also been widely adopted in the literature when studying the general second law \cite{MohseniSadjadi:2005ps,MohseniSadjadi:2010nu,Debnath:2011qga,Radicella:2010ss}.
Subsequently, the generalized second law can be further simplified to
\be\label{entropychange0}
\dot{S}_{t}=\frac{\dot{H}^{2}}{2H^{4}T}(f_{Q}+2Qf_{QQ})\geq0.
\ee
For a universe with positive temperature, Eq.~(\ref{entropychange0}) means
\be\label{Secondcondition0}
f_{Q}+2Qf_{QQ}\geq0.
\ee
In such a toy universe, the second law of generalized thermodynamics in the $f(Q)$ universe
is equivalent to a very simple inequality (\ref{Secondcondition0}).

Finally, let's look at some simple applications of inequality (\ref{Secondcondition0}).
For the STEGR model which is equivalent to GR,
the inequality (\ref{Secondcondition0}) can be reduced to $1>0$. This means that the general second law of thermodynamics is always satisfied in GR.
For the $f(Q)$ model that only retains the lowest-order correction, that is, $f(Q)=Q+\lambda\, Q^{2}$,
the inequality (\ref{Secondcondition0}) is reduced to $1-36\lambda H^{2}\geq0 $.
If the coefficient $\lambda<0$, the generalized second law always holds.
But if the coefficient $\lambda>0$, then the generalized second law requires $H^{2}<(36\lambda)^{-1}$,
that is, the expansion of the universe cannot be arbitrarily fast.
This example shows that the generalized second law can impose constraints on model parameters or the evolution of the universe.

\subsection{$P-V$ Phase Transition}\label{Sectransition}
Phase transitions and critical phenomena are fascinating topics in thermodynamics.
After having a preliminary understanding of the thermodynamic laws,
it is natural to ask whether there is a phase transition in the $f(Q)$ universe.
In the thermodynamic description of the $f(Q)$ universe, there are thermodynamic quantities such as pressure, temperature and volume.
Therefore, there may be a phase transition similar to that of a gas-liquid system.
In this subsection, we will analyze the $P-V$ phase transition of the $f(Q)$ universe and calculate its critical exponents.

From Eqs.~(\ref{temperature}) and (\ref{workdensity}), we can find the  equation of state $P=P(V,T)$ of the $f(Q)$ universe as
\be\label{EoS}
P=\frac{f}{16\pi}+\frac{f_{Q}}{2\pi R_{A}^{2}}+\frac{3f_{QQ}}{\pi R_{A}^{4}}+\left(\frac{f_{Q}}{2R_{A}}-\frac{6f_{QQ}}{R_{A}^{3}}\right)T,
\ee
where the horizon radius $R_{A}$ should be understood as $R_{A}(V)=(3V/4\pi)^{1/3}$ in this subsection.
For the STEGR model, the equation of state (\ref{EoS}) can be simplified to
\be\label{EoS0}
P=\frac{1}{8\pi R_{A}^{2}}+\frac{T}{2R_{A}}.
\ee
Given any non-negative temperature $T$, the pressure $P$ in Eq.~(\ref{EoS0}) is a monotonically decreasing function of volume $V$.
Therefore, there is no $P-V$ phase transition in GR.
This result is consistent with the research in Ref.~\cite{Abdusattar:2021wfv}.

Next, we examine the $f(Q)$ model that only retains the lowest-order correction to GR, that is, the model $f(Q)=Q+\lambda\, Q^{2}$.
In this model, the equation of state (\ref{EoS}) is reduced to
\be\label{EoS1}
P=\frac{1}{8{\pi}R_A^2}+\frac{T}{2R_{A}}+\frac{9\lambda}{4{\pi}R_A^4}-\frac{18{\lambda}T}{R_A^3}.
\ee
The necessary conditions for the $P-V$ phase transition are
\be\label{PTcondition}
\left(\frac{\partial P}{\partial V}\right)_T=\left(\frac{\partial^2P}{\partial V^{2}}\right)_T=0.
\ee
For the case of $\lambda>0$, Eq.~(\ref{PTcondition}) has the following physically reasonable solution
\be\label{criticalpoint}
T_c=\frac{\sqrt{3+2\sqrt{3}}}{36\pi\sqrt{\lambda}}~,~R_c=6\sqrt{(2\sqrt{3}-3)\lambda}~,~P_c=\frac{15+8\sqrt{3}}{5184\pi\lambda},
\ee
where $T_{c}$ is the critical temperature, $R_c$ is the horizon radius at the critical point, and $P_c$ is the pressure at the critical point.
The universal dimensionless quantity for this critical point is
\be
\frac{P_cR_c}{T_c}=1+\frac{\sqrt{3}}{6},
\ee
which is a constant and independent of the model parameter $\lambda$.
The above results show that there is a $P-V$ phase transition in the model $f(Q)=Q+\lambda\, Q^{2}$ with $\lambda>0$.
This can be seen more intuitively from the $P-R_{A}$ phase diagram.
\begin{figure}[h]
  \centering
  \includegraphics[width=0.7\textwidth]{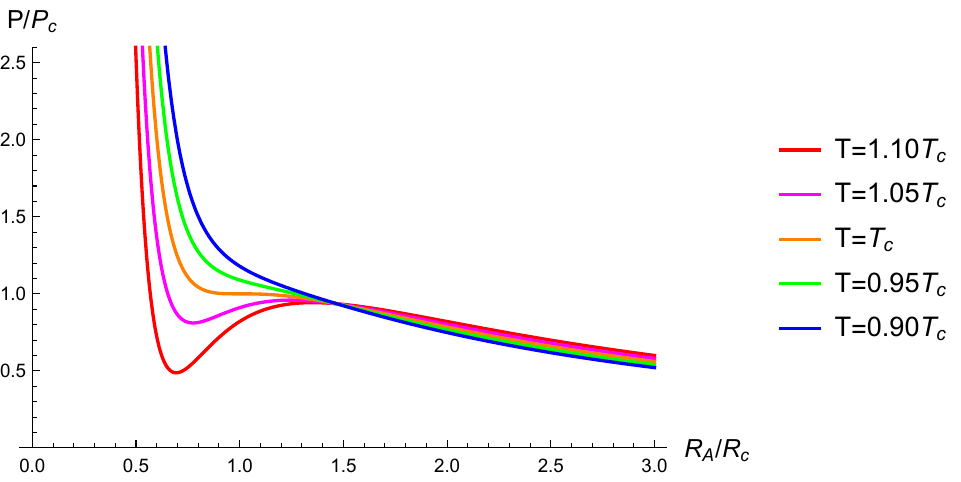}\\
  \caption{ Isotherms in the $P-R_{A}$ phase diagram}
  \label{PVfigure}
\end{figure}
Figure \ref{PVfigure} shows the isotherms near the critical temperature,
where warmer colors mean higher temperatures and cooler colors mean lower temperatures.
It can be seen that the shape of the isotherms is somewhat similar to those of  the van der Waals gas-liquid system,
some black hole systems \cite{Nam:2018ltb,Xu:2015rfa,Hendi:2017fxp,Hu:2018qsy,Wei:2020poh},
the special Horndeski universe \cite{Kong:2021dqd}  and the holographic dark energy model \cite{Cruz:2023xjp}.
However, there is a significant difference from the van der Waals system, that is,
the coexistence phase appears above the critical temperature $T>T_{c}$ in the $f(Q)$ universe,
whereas the coexistence phase appears below the critical temperature $T<T_{c}$ in the van der Waals system.
In addition, the critical temperature $T_{c}\rightarrow+\infty$ if $\lambda\rightarrow 0^{+}$,
which is consistent with the previous conclusion that there is no phase transition in the STEGR model.

Similar to the gas-liquid phase transition, we can define the critical exponents $\{\alpha,\beta,\gamma,\delta\}$ near the critical point as follows:
\bea
 C_{V}=T\left(\frac{{\partial}S}{{\partial}T}\right)_{V}~&\propto&~|t|^{-\alpha},\\
 \eta=\frac{V_l-V_s}{V_c}~&\propto&~|t|^{\beta},\\
 \kappa_T=-\frac{1}{V}\left(\frac{\partial V}{\partial P}\right)_T ~&\propto&~|t|^{-\gamma},\\
 p|_{t=0}~&\propto&~|v|^{\delta},
\eea
where
\be
 t=\frac{T-T_c}{T_c}~,~p=\frac{P-P_c}{P_c}~,~v=\frac{V-V_c}{V_c},
\ee
are the reduced temperature, pressure, and volume, respectively,
while $V_l$ and $V_s$ are two different volumes with equal pressure and Gibbs free energy.
In the language of gas-liquid systems, $C_{V}$ is the isovolumetric heat capacity,
$\kappa_{T}$ is the isothermal compressibility, and $\eta$ can be regarded as the order parameter.
In the following, we will calculate the four critical exponents and examine their scaling laws.

Since the entropy is a function only of volume and has nothing to do with temperature,
the isovolumetric heat capacity is zero, i.e.  $C_{V}=0$, which means the first critical exponent $\alpha=0$.
In order to obtain the other three critical ones, we expand the reduced pressure $p$ near the critical point
\begin{equation}\label{expansion}
    p=-\frac{8}{11}(2\sqrt{3}-1)t+\frac{8}{33}(3+5\sqrt{3})vt-\frac{4}{297}(5+\sqrt{3})v^3+...
\end{equation}
where "$...$" means unimportant higher-order terms.
Note that after we use the reduced thermodynamic variables,
the equation of state becomes independent of the model parameter $\lambda$.
From Eq.~(\ref{expansion}), one can easily find the isothermal compressibility and the pressure near the critical point as
\be
\kappa_{T} \propto \left(\frac{{\partial}p}{{\partial}v}\right)^{-1}{\propto}~t^{-1}~,~p|_{t=0} \propto v^{3},
\ee
which gives the critical exponents $\gamma=1$ and $\delta=3$.
To obtain the last critical exponent, we must solve for $v_s$ and $v_l$,
which are different reduced volumes with the same pressure and the same Gibbs free energy $G=U+PV-TS$.
Clearly, $v_l$ and $v_s$ satisfy the following two equations
\be\label{Veom}
p(v_s,t)=p(v_l,t)~~,~~\int vdp=\int_{v_{s}}^{v_l}v\left(\frac{\partial p}{\partial v}\right) \zd v=0.
\ee
From Eqs.~(\ref{expansion}) and (\ref{Veom}), one can get a nontrivial solution
\be\label{Vsol}
v_l=3\sqrt{2\sqrt{3}t}~,~v_s=-3\sqrt{2\sqrt{3}t}.
\ee
Therefore, $\eta=v_l-v_s\propto t^{1/2}$, and the last critical exponent $\beta=1/2$.

In summary, the four critical exponents are
\be\label{criticalexponent}
\alpha=0~,~\beta=\frac{1}{2}~,~\gamma=1~,~\delta=3.
\ee
Somewhat surprisingly, these critical exponents are exactly the same as those in the mean field theory.
So of course they obey the usual scaling laws
\bea\label{scalinglaws}
& &\nonumber
\alpha+2\beta+\gamma=2,~\alpha+\beta(1+\delta)=2,
\\ & &\gamma(1+\delta)=(2-\alpha)(\delta-1),~\gamma=\beta(\delta-1).
\eea
In subsection \ref{Secfirstlaw},
we showed that the lowest-order correction to GR by the $f(Q)$ model leads to a logarithmic correction to the entropy,
which is consistent with the lowest-order correction in loop quantum gravity.
In this subsection, we show that the lowest-order correction to GR by the $f(Q)$ model can yield the $P-V$ phase transition,
and the critical exponents are exactly the same as those in the mean field theory.
These unexpected coincidences seem to hint that the correction brought by the $f(Q)$ model may emerge from some quantum mechanism, and its microscopic statistics satisfy the mean field theory in the lowest-order approximation.
Of course, we have no evidence to support this idea so far, and it could just be a coincidence.

\section{Thermodynamic implication of nontrivial connections}\label{Secnontrivial}

In section \ref{Sectrivial}, we analyzed the thermodynamics of the $f(Q)$ universe with trivial connection
from the perspectives of the first law, generalized second law and $P-V$ phase transition.
As mentioned in section \ref{SecfQ}, there are flat universes with nontrivial connection in the $f(Q)$ model.
In this section, we will explore the thermodynamic implications of these nontrivial connections in the $f(Q)$ flat universe.

In fact, it is unprecedented to discuss how different connections affect the thermodynamics of the universe in the same gravity model.
Because in most gravity models, such as $f(R)$ and $f(T)$ models, the connection that satisfies cosmological symmetry is unique.
Without different connections, naturally there would be no discussion of the thermodynamic effects of different connections.
As a result, our exploration in this section is somewhat pioneering.

In the $f(Q)$ flat universe, the connection that obeys cosmological symmetry has three branch solutions, as shown in section \ref{SecfQ}.
For the branch 1 solution, one can find that
the background equations in the branch 1 universe are exactly the same as those in the universe with trivial connection.
Therefore, the branch 1 universe should also be classified as a universe with trivial connection.
To analyze the thermodynamic effects of nontrivial connections, we should focus on branch 2 and branch 3 universes.

For the branch 2 universe, following the process in subsection \ref{Secfirstlaw}, we obtain the energy and work density as
\bea
& &\label{energy2}
E=\frac{R_Af_Q}{2}+\frac{R_A^3}{12}(f-Qf_Q)+\frac{R_{A}^{3}}{4}\gamma\dot{f}_{Q},
\\ & &\label{workdensity2}
W=\frac{1}{8\pi}\Big[f_Q(\dot{H}+3H^2)+\frac{1}{2}(f-Qf_Q)+H\dot{f}_{Q}\Big].
\eea
After identifying the internal energy $U=-E$ and the thermodynamic pressure $P=W$,
the first law of thermodynamic (\ref{firstlaw}) can be established as long as
\bea\label{deltaS2}
\zd S=\bigg\{2\pi R_{A}f_{Q}+3\pi f_{QQ}\Big[18\gamma-\frac{8}{R_{A}}-9R_{A}\gamma^{2}
+(3R_{A}\gamma-2)(R_{A}\ddot{\gamma}+3\dot{\gamma})\frac{R_{A}}{\dot{R}_{A}}\Big]\bigg\}\,\zd R_{A},~~
\eea
where $Q=-6R_{A}^{-2}+9\gamma R_{A}^{-1}+3\dot{\gamma}$ and $\gamma$ satisfies the following equation due to Eq.~(\ref{beom3})
\be\label{gammaeom2}
\gamma(\ddot{f}_{Q}+3H\dot{f}_{Q})=0.
\ee
At first glance, the entropy in Eq.~(\ref{deltaS2}) is not-integrable unless $\gamma=0$ or $f_{Q}=\text{const}$.
To see this more clearly, let us again consider the simple model $f(Q)=Q+\lambda Q^{2}$.
In this model, the differential of entropy (\ref{deltaS2}) can be simplified to
\be\label{dentropy2}
\zd S=\zd\left(\frac{A}{4}-36\pi\lambda\ln\frac{A}{A_0}\right)+\zd \hat{S},
\ee
where
\be\label{deltaS02}
\zd \hat{S}=6\pi\lambda\left\{24\gamma\dot{R}_{A}+R_{A}
\left[(2\dot{\gamma}-9\gamma^{2})\dot{R}_{A}+(3R_{A}\gamma-2)(R_{A}\ddot{\gamma}+3\dot{\gamma})\right]\right\}\zd t.
\ee
If $\lambda=0$  (equivalent to GR) or $\gamma=0$ (trivial connection), then $\zd \hat{S}=0$,
so the entropy $S$ in Eq.~(\ref{dentropy2}) can be integrated as a function of the horizon area.
If $\lambda\neq 0$ and $\gamma\neq0$, then $\gamma$ should satisfy the following equation due to Eq.~(\ref{gammaeom2})
\be\label{gammaeom02}
\dddot{\gamma}+\frac{6}{R_{A}}\ddot{\gamma}+\frac{3(3-2\dot{R}_{A})}{R_{A}^{2}}\dot{\gamma}
-\frac{3(3\dot{R}_{A}-2\dot{R}_{A}^{2}+R_{A}\ddot{R}_{A})}{R_{A}^{3}}\gamma+\frac{4(3\dot{R}_{A}-3\dot{R}_{A}^{2}+R_{A}\ddot{R}_{A})}{R_{A}^{4}}=0.~~
\ee
Since the dependence of $\gamma$ on $\{R_{A}, \dot{R}_{A}, \ddot{R}_{A}\}$ is realized through a third-order differential equation,
it is impossible to eliminate $\gamma$ in Eq.~(\ref{deltaS02}) by Eq.~(\ref{gammaeom02}).
This means that $\hat{S}$ cannot be integrated as a function of the horizon radius $R_{A}$.
For the same reason, the entropy in Eq.~(\ref{dentropy2}) cannot be integrated into a function only about $R_{A}$ and $\dot{R}_{A}$,
that is, it cannot be expressed as a function $S=S(V,T)$ only about volume and temperature.
A more rigorous proof can be found in Appendix \ref{Appintegrability}.
From Eq.~(\ref{deltaS2}) we can expect the same thing to happen in other $f(Q)$ models.

The situation of the branch 3 universe is almost the same as that of the branch 2 universe.
In the branch 3 universe, the energy and work density are
\bea
& &\label{energy3}
E=\frac{R_Af_Q}{2}+\frac{R_A^3}{12}(f-Qf_Q)-\frac{R_{A}^{3}}{4}\gamma\dot{f}_{Q},
\\ & &\label{workdensity3}
W=\frac{1}{8\pi}\Big[f_Q(\dot{H}+3H^2)+\frac{1}{2}(f-Qf_Q)+(H-\gamma)\dot{f}_{Q}\Big].
\eea
After identifying the internal energy $U=-E$ and the thermodynamic pressure $P=W$,
the first law of thermodynamic (\ref{firstlaw}) gives
\bea\label{deltaS3}
\zd S=\bigg\{2\pi R_{A}f_{Q}+3\pi f_{QQ}\Big[-\frac{8}{R_{A}}+2\gamma+3R_{A}\gamma^{2}
-(2+R_{A}\gamma)(R_{A}\ddot{\gamma}+3\dot{\gamma})\frac{R_{A}}{\dot{R}_{A}}\Big]\bigg\}\,\zd R_{A},~~
\eea
where $Q=-6R_{A}^{-2}+9\gamma R_{A}^{-1}+3\dot{\gamma}$ and $\gamma\ddot{f}_{Q}+(2\dot{\gamma}+5H\gamma)\dot{f}_{Q}=0$.
Once again, the entropy in Eq.~(\ref{deltaS3}) appears to be non-integrable unless $\gamma=0$ or $f_{Q}=\text{const}$.
This can be seen more clearly in a simple example of $f(Q)=Q+\lambda Q^{2}$.
In this model, the differential of entropy (\ref{deltaS2}) can be simplified to
\be\label{dentropy3}
\zd S=\zd\left(\frac{A}{4}-36\pi\lambda\ln\frac{A}{A_0}\right)+\zd \hat{S},
\ee
with
\be\label{deltaS03}
\zd \hat{S}=6\pi\lambda\left\{8\gamma\dot{R}_{A}+R_{A}
\left[(2\dot{\gamma}+3\gamma^{2})\dot{R}_{A}-(2+R_{A}\gamma)(R_{A}\ddot{\gamma}+3\dot{\gamma})\right]\right\}\zd t,
\ee
where $\gamma$ satisfies the equation
\bea\label{gammaeom03}
& &\nonumber
\gamma\dddot{\gamma}+2\dot{\gamma}\ddot{\gamma}+\frac{2}{R_{A}}(4\gamma\ddot{\gamma}+3\dot{\gamma}^{2})+\frac{3(5-4\dot{R}_{A})}{R_{A}^{2}}\gamma\dot{\gamma}
\\ & &\quad\quad
-\frac{3(5\dot{R}_{A}-2\dot{R}_{A}^{2}+R_{A}\ddot{R}_{A})}{R_{A}^{3}}\gamma^{2}
+\frac{8\dot{R}_{A}}{R_{A}^{3}}\dot{\gamma}+\frac{4(5\dot{R}_{A}-3\dot{R}_{A}^{2}+R_{A}\ddot{R}_{A})}{R_{A}^{4}}\gamma=0.
\eea
For the same reasons as for the branch 2 universe,
the entropy $S$ in Eq.~(\ref{dentropy3}) cannot be integrated as a function of $R_{A}$ and $\dot{R}_{A}$ unless $\gamma = 0$ or $\lambda = 0$.

In summary, in the $f(Q)$ flat universe with nontrivial connection,
the first law of thermodynamic gives
\bea\label{deltaS4}
\zd S=\zd S(A)+\zd \hat{S}~~\text{with}~~\zd \hat{S}=\mF(\gamma,\dot{\gamma},\ddot{\gamma},R_{A},\dot{R}_{A})\zd t,
\eea
where $\gamma$ depends on $\{R_{A}, \dot{R}_{A}, \ddot{R}_{A}\}$ through a third-order differential equation.
Consequently, the entropy cannot be integrated as a function of the horizon radius and the horizon temperature in general.
The change of entropy depends on more specific process of cosmic evolutions.
Following the discussions in Refs.~\cite{Cai:2006rs,Akbar:2006mq,Miao:2011ki,Eling:2006aw},
the new term $\zd\hat{S}$ can be interpreted as the entropy production term.
This reveals that the horizon thermodynamics in the $f(Q)$ universe with nontrivial connection is non-equilibrium thermodynamics
\footnote{
Just as the $f(R)$ and $f(T)$ universes can be described by equilibrium thermodynamic \cite{Bamba:2009id,Bamba:2011pz,Bamba:2011jq},
the $f(Q)$ universe with nontrivial connection can also be described by equilibrium thermodynamics.
However, the realization of the equilibrium description
can be independent of the gravity model and spacetime connection (see Appendix \ref{Appequilibrium}),
so we do not adopt this perspective in this paper.
}.
This is a significant feature of nontrivial connections from the perspective of thermodynamics.

\section{Conclusion}\label{Conclusion}

In this paper, we have analyzed the thermodynamics of apparent horizon in the $f(Q)$ universe with trivial and nontrivial connections.
For the case of the trivial connection, first,
we have studied its first law and showed that the first law in thermal equilibrium holds for any $f(Q)$ models.
We have also obtained the area law of entropy for the most general $f(Q)$ model
and showed that the lowest-order and second-lowest-order corrections to the entropy have the same form as those in loop quantum gravity.
Then, we briefly investigated the generalized second law,
which asserts that the entropy of the horizon plus the entropy of matter within the horizon does not decrease.
After assuming that all matter is in thermal equilibrium with the horizon,
we derived a simple inequality (\ref{Secondcondition0}) from the generalized second law that holds for any $f(Q)$ models.
Finally, we have studied the $P-V$ phase transition by analogy with the gas-liquid system.
We have shown that even the $f(Q)$ model that retains only the lowest-order correction to GR can lead to the $P-V$ phase transition that does not exist in GR.
We have calculated the critical exponents and found that they are exactly the same as those in mean field theory.
For the case of nontrivial connections,
we have examined the thermodynamic effect of nontrivial connections from the perspective of the first law.
We have proved that the change in entropy cannot be determined entirely by the change in volume and temperature.
This means that the horizon thermodynamics of the $f(Q)$ universe with nontrivial connection is non-equilibrium.
This may be the first exploration of how different connections affect the thermodynamics of the universe in the same gravity model.

\subsection*{Acknowledgements}
This work is supported in part by
the National Key Research and Development Program of China under Grant No. 2020YFC2201501
and the National Natural Science Foundation of China (NSFC) under Grant No. 12347103 and 12205063.

\appendix

\section{The condition for $S=S(V,T)$}\label{Appintegrability}

In this appendix, we will discuss the condition for the integrability of entropy,
and further prove that the entropies in Eqs.~(\ref{deltaS02}) and (\ref{deltaS03}) in the main text are not integrable in the thermodynamic sense.

In the thermodynamics of the horizon, sometimes entropy can be integrated as a function of the horizon radius.
This is a remarkable thermodynamic property that is not always possible.
Once this fails, some might conclude that entropy is non-integrable.
But we think it is too hasty to draw such a conclusion.
Because it is also possible that entropy can be integrated as $S=S(R_{A},\dot{R}_{A})$,
so that entropy is a function of volume $V=\frac{4}{3}\pi R_{A}^{3}$ and temperature (\ref{temperature}).
This means that the change of entropy only depends on the initial and final states of the system, so such entropy is integrable.

If entropy can be integrated as a function of volume and temperature, that is, $S=S(R_{A},\dot{R}_{A})$, then the differential of entropy is
\be
\zd S= \frac{\partial S}{\partial R_{A}}\zd R_{A}+\frac{\partial S}{\partial \dot{R}_{A}}\zd \dot{R}_{A}
=\left(\frac{\partial S}{\partial R_{A}}\dot{R}_{A}+\frac{\partial S}{\partial \dot{R}_{A}}\ddot{R}_{A}\right)\zd t.
\ee
As as result, the differential of entropy must have the form
\be\label{appA1}
\zd S=\left[f_{1}(R_{A},\dot{R}_{A})+f_{2}(R_{A},\dot{R}_{A})\ddot{R}_{A}\right]\zd t.
\ee
Since the order of partial derivatives of smooth functions is commutative, that is
\be
\frac{\partial^{2} S}{\partial R_{A}\partial \dot{R}_{A}}=\frac{\partial^{2} S}{\partial \dot{R}_{A}\partial {R}_{A}},
\ee
$f_{1}(R_{A},\dot{R}_{A})$ and $f_{2}(R_{A},\dot{R}_{A})$ must satisfy the following condition:
\be\label{appA2}
\dot{R}_{A}\frac{\partial f_1}{\partial \dot{R}_{A}}-f_{1}=\dot{R}_{A}^{2}\frac{\partial f_2}{\partial {R}_{A}}.
\ee
Therefore, if entropy is integrable in the thermodynamic sense,
the differential of entropy should be consistent with the form of Eq.~(\ref{appA1}) and satisfy the condition (\ref{appA2}).

However, the differential of entropy in Eq.~(\ref{deltaS02}) cannot even be consistent with the form of Eq.~(\ref{appA1}),
because the dependence of $\gamma$ on $\{R_{A}, \dot{R}_{A}, \ddot{R}_{A}\}$
is implemented through a differential equation rather than a algebraic equation.
Consequently, the differential of entropy in Eq.~(\ref{deltaS02}) is not integrable at least in the thermodynamic sense.
For the same reason, the differential entropy in Eq.~(\ref{deltaS03}) is also not integrable in the thermodynamic sense.

\section{Thermal equilibrium description of horizon thermodynamics}\label{Appequilibrium}

In this appendix, we will show that the horizon in a flat universe can always be described by equilibrium thermodynamics.

In a flat universe, the background equations can be written as
\bea
& & 3H^{2}=8\pi(\rho_{m}+\rho_{MG}),\\
& &2\dot{H}+3H^{2}=-8\pi (p_{m}+p_{MG}),
\eea
where $\rho_{MG}$ and $p_{MG}$ are the effective energy density and effective pressure resulting from the modified gravity.
With the internal energy $U=-(\rho_{m}+\rho_{MG})V$ and the thermodynamic pressure $P=\frac{1}{2}(\rho_{m}+\rho_{MG}-p_{m}-p_{MG})$,
the first law of thermodynamics (\ref{firstlaw}) holds,
where the entropy $S$ is always equal to $A/4$, regardless of the specific gravity model.
Since the above derivation depends neither on a specific gravity model nor on spacetime connection,
it is applicable to any flat universes with cosmological symmetry in any gravity models.
The details of the gravity model are contained in $\rho_{MG}$ and $p_{MG}$.

In this perspective, the characteristics of the $f(Q)$ model
and the distinction between trivial and nontrivial connections are difficult to capture.
This is why we did not adopt the equilibrium description in this paper.

{}


\begin{thebibliography}{}




\bibitem{Bardeen:1973gs}
J.~M.~Bardeen, B.~Carter and S.~W.~Hawking,
Commun. Math. Phys. \textbf{31}, 161-170 (1973)
doi:10.1007/BF01645742

\bibitem{Bekenstein:1973ur}
J.~D.~Bekenstein,
Phys. Rev. D \textbf{7}, 2333-2346 (1973)
doi:10.1103/PhysRevD.7.2333

\bibitem{Hawking:1975vcx}
S.~W.~Hawking,
Commun. Math. Phys. \textbf{43}, 199-220 (1975)
[erratum: Commun. Math. Phys. \textbf{46}, 206 (1976)]
doi:10.1007/BF02345020


\bibitem{Jacobson:1995ab}
T.~Jacobson,
Phys. Rev. Lett. \textbf{75}, 1260-1263 (1995)
doi:10.1103/PhysRevLett.75.1260
[arXiv:gr-qc/9504004 [gr-qc]].

\bibitem{Jacobson:2015hqa}
T.~Jacobson,
Phys. Rev. Lett. \textbf{116}, no.20, 201101 (2016)
doi:10.1103/PhysRevLett.116.201101
[arXiv:1505.04753 [gr-qc]].

\bibitem{Padmanabhan:2009vy}
T.~Padmanabhan,
Rept. Prog. Phys. \textbf{73}, 046901 (2010)
doi:10.1088/0034-4885/73/4/046901
[arXiv:0911.5004 [gr-qc]].

\bibitem{Wald:1999vt}
R.~M.~Wald,
Living Rev. Rel. \textbf{4}, 6 (2001)
doi:10.12942/lrr-2001-6
[arXiv:gr-qc/9912119 [gr-qc]].




\bibitem{Kubiznak:2016qmn}
D.~Kubiznak, R.~B.~Mann and M.~Teo,
Class. Quant. Grav. \textbf{34}, no.6, 063001 (2017)
doi:10.1088/1361-6382/aa5c69
[arXiv:1608.06147 [hep-th]].

\bibitem{Hawking:1982dh}
S.~W.~Hawking and D.~N.~Page,
Commun. Math. Phys. \textbf{87}, 577 (1983)
doi:10.1007/BF01208266

\bibitem{Altamirano:2014tva}
N.~Altamirano, D.~Kubiznak, R.~B.~Mann and Z.~Sherkatghanad,
Galaxies \textbf{2}, 89-159 (2014)
doi:10.3390/galaxies2010089
[arXiv:1401.2586 [hep-th]].




\bibitem{Cai:2001dz}
R.~G.~Cai,
Phys. Rev. D \textbf{65}, 084014 (2002)
doi:10.1103/PhysRevD.65.084014
[arXiv:hep-th/0109133 [hep-th]].

\bibitem{Cai:2003kt}
R.~G.~Cai,
Phys. Lett. B \textbf{582}, 237-242 (2004)
doi:10.1016/j.physletb.2004.01.015
[arXiv:hep-th/0311240 [hep-th]].


\bibitem{Akbar:2006mq}
M.~Akbar and R.~G.~Cai,
Phys. Lett. B \textbf{648}, 243-248 (2007)
doi:10.1016/j.physletb.2007.03.005
[arXiv:gr-qc/0612089 [gr-qc]].

\bibitem{Miao:2011ki}
R.~X.~Miao, M.~Li and Y.~G.~Miao,
JCAP \textbf{11}, 033 (2011)
doi:10.1088/1475-7516/2011/11/033
[arXiv:1107.0515 [hep-th]].





\bibitem{Padmanabhan:2003gd}
T.~Padmanabhan,
Phys. Rept. \textbf{406}, 49-125 (2005)
doi:10.1016/j.physrep.2004.10.003
[arXiv:gr-qc/0311036 [gr-qc]].

\bibitem{Cai:2005ra}
R.~G.~Cai and S.~P.~Kim,
JHEP \textbf{02}, 050 (2005)
doi:10.1088/1126-6708/2005/02/050
[arXiv:hep-th/0501055 [hep-th]].





\bibitem{Gibbons:1977mu}
G.~W.~Gibbons and S.~W.~Hawking,
Phys. Rev. D \textbf{15}, 2738-2751 (1977)
doi:10.1103/PhysRevD.15.2738

\bibitem{Cai:2008gw}
R.~G.~Cai, L.~M.~Cao and Y.~P.~Hu,
Class. Quant. Grav. \textbf{26}, 155018 (2009)
doi:10.1088/0264-9381/26/15/155018
[arXiv:0809.1554 [hep-th]].

\bibitem{DiCriscienzo:2009kun}
R.~Di Criscienzo, S.~A.~Hayward, M.~Nadalini, L.~Vanzo and S.~Zerbini,
Class. Quant. Grav. \textbf{27}, 015006 (2010)
doi:10.1088/0264-9381/27/1/015006
[arXiv:0906.1725 [gr-qc]].

\bibitem{Zhu:2009wa}
T.~Zhu, J.~R.~Ren and D.~Singleton,
Int. J. Mod. Phys. D \textbf{19}, 159-169 (2010)
doi:10.1142/S0218271810016336
[arXiv:0902.2542 [hep-th]].

\bibitem{Akbar:2006kj}
M.~Akbar and R.~G.~Cai,
Phys. Rev. D \textbf{75}, 084003 (2007)
doi:10.1103/PhysRevD.75.084003
[arXiv:hep-th/0609128 [hep-th]].

\bibitem{Cai:2006rs}
R.~G.~Cai and L.~M.~Cao,
Phys. Rev. D \textbf{75}, 064008 (2007)
doi:10.1103/PhysRevD.75.064008
[arXiv:gr-qc/0611071 [gr-qc]].


\bibitem{Kong:2021dqd}
S.~B.~Kong, H.~Abdusattar, Y.~Yin, H.~Zhang and Y.~P.~Hu,
Eur. Phys. J. C \textbf{82}, no.11, 1047 (2022)
doi:10.1140/epjc/s10052-022-10976-9
[arXiv:2108.09411 [gr-qc]].









\bibitem{Akbar:2006er}
M.~Akbar and R.~G.~Cai,
Phys. Lett. B \textbf{635}, 7-10 (2006)
doi:10.1016/j.physletb.2006.02.035
[arXiv:hep-th/0602156 [hep-th]].

\bibitem{Gong:2007md}
Y.~Gong and A.~Wang,
Phys. Rev. Lett. \textbf{99}, 211301 (2007)
doi:10.1103/PhysRevLett.99.211301
[arXiv:0704.0793 [hep-th]].

\bibitem{Bamba:2009id}
K.~Bamba, C.~Q.~Geng and S.~Tsujikawa,
Phys. Lett. B \textbf{688}, 101-109 (2010)
doi:10.1016/j.physletb.2010.03.070
[arXiv:0909.2159 [gr-qc]].

\bibitem{Bamba:2011pz}
K.~Bamba and C.~Q.~Geng,
JCAP \textbf{11}, 008 (2011)
doi:10.1088/1475-7516/2011/11/008
[arXiv:1109.1694 [gr-qc]].

\bibitem{Bamba:2009ay}
K.~Bamba and C.~Q.~Geng,
Phys. Lett. B \textbf{679}, 282-287 (2009)
doi:10.1016/j.physletb.2009.07.039
[arXiv:0901.1509 [hep-th]].


\bibitem{Bamba:2009gq}
K.~Bamba, C.~Q.~Geng, S.~Nojiri and S.~D.~Odintsov,
EPL \textbf{89}, no.5, 50003 (2010)
doi:10.1209/0295-5075/89/50003
[arXiv:0909.4397 [hep-th]].

\bibitem{Bamba:2011jq}
K.~Bamba, C.~Q.~Geng and S.~Tsujikawa,
Int. J. Mod. Phys. D \textbf{20}, 1363-1371 (2011)
doi:10.1142/S0218271811019542
[arXiv:1101.3628 [gr-qc]].

\bibitem{Bamba:2010kf}
K.~Bamba and C.~Q.~Geng,
JCAP \textbf{06}, 014 (2010)
doi:10.1088/1475-7516/2010/06/014
[arXiv:1005.5234 [gr-qc]].












\bibitem{Nester:1998mp}
J.~M.~Nester and H.~J.~Yo,
Chin. J. Phys. \textbf{37}, 113 (1999)
[arXiv:gr-qc/9809049 [gr-qc]].

\bibitem{BeltranJimenez:2019esp}
J.~Beltr\'an Jim\'enez, L.~Heisenberg and T.~S.~Koivisto,
Universe \textbf{5}, no.7, 173 (2019)
doi:10.3390/universe5070173
[arXiv:1903.06830 [hep-th]].

\bibitem{Capozziello:2022zzh}
S.~Capozziello, V.~De Falco and C.~Ferrara,
Eur. Phys. J. C \textbf{82}, no.10, 865 (2022)
doi:10.1140/epjc/s10052-022-10823-x
[arXiv:2208.03011 [gr-qc]].





\bibitem{BeltranJimenez:2017tkd}
J.~Beltr\'an Jim\'enez, L.~Heisenberg and T.~Koivisto,
Phys. Rev. D \textbf{98}, no.4, 044048 (2018)
doi:10.1103/PhysRevD.98.044048
[arXiv:1710.03116 [gr-qc]].

\bibitem{Heisenberg:2023lru}
L.~Heisenberg,
Phys. Rept. \textbf{1066}, 1-78 (2024)
doi:10.1016/j.physrep.2024.02.001
[arXiv:2309.15958 [gr-qc]].


\bibitem{BeltranJimenez:2019tme}
J.~Beltr\'an Jim\'enez, L.~Heisenberg, T.~S.~Koivisto and S.~Pekar,
Phys. Rev. D \textbf{101}, no.10, 103507 (2020)
doi:10.1103/PhysRevD.101.103507
[arXiv:1906.10027 [gr-qc]].


\bibitem{Atayde:2021pgb}
L.~Atayde and N.~Frusciante,
Phys. Rev. D \textbf{104}, no.6, 064052 (2021)
doi:10.1103/PhysRevD.104.064052
[arXiv:2108.10832 [astro-ph.CO]].


\bibitem{Oliveros:2023mwl}
A.~Oliveros and M.~A.~Acero,
Int. J. Mod. Phys. D \textbf{33}, no.01, 2450004 (2024)
doi:10.1142/S0218271824500044
[arXiv:2311.01857 [astro-ph.CO]].

\bibitem{Khyllep:2022spx}
W.~Khyllep, J.~Dutta, E.~N.~Saridakis and K.~Yesmakhanova,
Phys. Rev. D \textbf{107}, no.4, 044022 (2023)
doi:10.1103/PhysRevD.107.044022
[arXiv:2207.02610 [gr-qc]].

\bibitem{Zhao:2021zab}
D.~Zhao,
Eur. Phys. J. C \textbf{82}, no.4, 303 (2022)
doi:10.1140/epjc/s10052-022-10266-4
[arXiv:2104.02483 [gr-qc]].

\bibitem{Hu:2023ndc}
K.~Hu, T.~Paul and T.~Qiu,
Sci. China Phys. Mech. Astron. \textbf{67}, no.2, 220413 (2024)
doi:10.1007/s11433-023-2275-0
[arXiv:2308.00647 [hep-th]].

\bibitem{Hohmann:2021ast}
M.~Hohmann,
Phys. Rev. D \textbf{104}, no.12, 124077 (2021)
doi:10.1103/PhysRevD.104.124077
[arXiv:2109.01525 [gr-qc]].

\bibitem{Heisenberg:2022mbo}
L.~Heisenberg, M.~Hohmann and S.~Kuhn,
Eur. Phys. J. C \textbf{83}, no.4, 315 (2023)
doi:10.1140/epjc/s10052-023-11462-6
[arXiv:2212.14324 [gr-qc]].

\bibitem{Hohmann:2020zre}
M.~Hohmann,
Int. J. Geom. Meth. Mod. Phys. \textbf{18}, no.supp01, 2140005 (2021)
doi:10.1142/S0219887821400053
[arXiv:2008.12186 [gr-qc]].


\bibitem{Gomes:2023hyk}
D.~A.~Gomes, J.~Beltr\'an Jim\'enez and T.~S.~Koivisto,
JCAP \textbf{12}, 010 (2023)
doi:10.1088/1475-7516/2023/12/010
[arXiv:2309.08554 [gr-qc]].



\bibitem{Dimakis:2022rkd}
N.~Dimakis, A.~Paliathanasis, M.~Roumeliotis and T.~Christodoulakis,
Phys. Rev. D \textbf{106}, no.4, 043509 (2022)
doi:10.1103/PhysRevD.106.043509
[arXiv:2205.04680 [gr-qc]].

\bibitem{Shi:2023kvu}
J.~Shi,
Eur. Phys. J. C \textbf{83}, no.10, 951 (2023)
doi:10.1140/epjc/s10052-023-12139-w
[arXiv:2307.08103 [gr-qc]].


\bibitem{Gomes:2023tur}
D.~A.~Gomes, J.~Beltr\'an Jim\'enez, A.~J.~Cano and T.~S.~Koivisto,
Phys. Rev. Lett. \textbf{132}, no.14, 141401 (2024)
doi:10.1103/PhysRevLett.132.141401
[arXiv:2311.04201 [gr-qc]].


\bibitem{Rao:2023nip}
H.~Rao, C.~Liu and C.~Q.~Geng,
Phys. Lett. B \textbf{850}, 138497 (2024)
doi:10.1016/j.physletb.2024.138497
[arXiv:2311.06600 [gr-qc]].












\bibitem{Bak:1999hd}
D.~Bak and S.~J.~Rey,
Class. Quant. Grav. \textbf{17}, L83 (2000)
doi:10.1088/0264-9381/17/15/101
[arXiv:hep-th/9902173 [hep-th]].

\bibitem{Hayward:1998ee}
S.~A.~Hayward, S.~Mukohyama and M.~C.~Ashworth,
Phys. Lett. A \textbf{256}, 347-350 (1999)
doi:10.1016/S0375-9601(99)00225-X
[arXiv:gr-qc/9810006 [gr-qc]].

\bibitem{Hayward:1997jp}
S.~A.~Hayward,
Class. Quant. Grav. \textbf{15}, 3147-3162 (1998)
doi:10.1088/0264-9381/15/10/017
[arXiv:gr-qc/9710089 [gr-qc]].


\bibitem{Meissner:2004ju}
K.~A.~Meissner,
Class. Quant. Grav. \textbf{21}, 5245-5252 (2004)
doi:10.1088/0264-9381/21/22/015
[arXiv:gr-qc/0407052 [gr-qc]].

\bibitem{Ghosh:2004rq}
A.~Ghosh and P.~Mitra,
Phys. Rev. D \textbf{71}, 027502 (2005)
doi:10.1103/PhysRevD.71.027502
[arXiv:gr-qc/0401070 [gr-qc]].


\bibitem{Chatterjee:2003uv}
A.~Chatterjee and P.~Majumdar,
Phys. Rev. Lett. \textbf{92}, 141301 (2004)
doi:10.1103/PhysRevLett.92.141301
[arXiv:gr-qc/0309026 [gr-qc]].


\bibitem{Banerjee:2009tz}
R.~Banerjee and S.~K.~Modak,
JHEP \textbf{05}, 063 (2009)
doi:10.1088/1126-6708/2009/05/063
[arXiv:0903.3321 [hep-th]].


\bibitem{Modak:2008tg}
S.~K.~Modak,
Phys. Lett. B \textbf{671}, 167-173 (2009)
doi:10.1016/j.physletb.2008.11.043
[arXiv:0807.0959 [hep-th]].


\bibitem{MohseniSadjadi:2010nu}
H.~Mohseni Sadjadi and M.~Jamil,
EPL \textbf{92}, no.6, 69001 (2010)
doi:10.1209/0295-5075/92/69001
[arXiv:1002.3588 [gr-qc]].



\bibitem{Das:2007mj}
S.~Das, S.~Shankaranarayanan and S.~Sur,
Phys. Rev. D \textbf{77}, 064013 (2008)
doi:10.1103/PhysRevD.77.064013
[arXiv:0705.2070 [gr-qc]].

\bibitem{Radicella:2010ss}
N.~Radicella and D.~Pavon,
Phys. Lett. B \textbf{691}, 121-126 (2010)
doi:10.1016/j.physletb.2010.06.019
[arXiv:1006.3745 [gr-qc]].

\bibitem{Sheykhi:2011egx}
A.~Sheykhi and M.~Jamil,
Gen. Rel. Grav. \textbf{43}, 2661-2672 (2011)
doi:10.1007/s10714-011-1190-x
[arXiv:1011.0134 [physics.gen-ph]].





\bibitem{Davies:1987ti}
P.~C.~W.~Davies,
Class. Quant. Grav. \textbf{4}, L225 (1987)
doi:10.1088/0264-9381/4/6/006

\bibitem{Clifton:2007tn}
T.~Clifton and J.~D.~Barrow,
Phys. Rev. D \textbf{75}, 043515 (2007)
doi:10.1103/PhysRevD.75.043515
[arXiv:gr-qc/0701070 [gr-qc]].

\bibitem{MohseniSadjadi:2005ps}
H.~Mohseni Sadjadi,
Phys. Rev. D \textbf{73}, 063525 (2006)
doi:10.1103/PhysRevD.73.063525
[arXiv:gr-qc/0512140 [gr-qc]].


\bibitem{Debnath:2011qga}
U.~Debnath, S.~Chattopadhyay, I.~Hussain, M.~Jamil and R.~Myrzakulov,
Eur. Phys. J. C \textbf{72}, 1875 (2012)
doi:10.1140/epjc/s10052-012-1875-7
[arXiv:1111.3858 [gr-qc]].

\bibitem{Abdusattar:2021wfv}
H.~Abdusattar, S.~B.~Kong, W.~L.~You, H.~Zhang and Y.~P.~Hu,
JHEP \textbf{12}, 168 (2022)
doi:10.1007/JHEP12(2022)168
[arXiv:2108.09407 [gr-qc]].






\bibitem{Nam:2018ltb}
C.~H.~Nam,
Eur. Phys. J. C \textbf{78}, no.12, 1016 (2018)
doi:10.1140/epjc/s10052-018-6498-1

\bibitem{Xu:2015rfa}
J.~Xu, L.~M.~Cao and Y.~P.~Hu,
Phys. Rev. D \textbf{91}, no.12, 124033 (2015)
doi:10.1103/PhysRevD.91.124033
[arXiv:1506.03578 [gr-qc]].

\bibitem{Hendi:2017fxp}
S.~H.~Hendi, R.~B.~Mann, S.~Panahiyan and B.~Eslam Panah,
Phys. Rev. D \textbf{95}, no.2, 021501 (2017)
doi:10.1103/PhysRevD.95.021501
[arXiv:1702.00432 [gr-qc]].

\bibitem{Hu:2018qsy}
Y.~P.~Hu, H.~A.~Zeng, Z.~M.~Jiang and H.~Zhang,
Phys. Rev. D \textbf{100}, no.8, 084004 (2019)
doi:10.1103/PhysRevD.100.084004
[arXiv:1812.09938 [gr-qc]].


\bibitem{Wei:2020poh}
S.~W.~Wei and Y.~X.~Liu,
Phys. Rev. D \textbf{101}, no.10, 104018 (2020)
doi:10.1103/PhysRevD.101.104018
[arXiv:2003.14275 [gr-qc]].






\bibitem{Cruz:2023xjp}
M.~Cruz, S.~Lepe and J.~Saavedra,
[arXiv:2312.14257 [gr-qc]].








\bibitem{Eling:2006aw}
C.~Eling, R.~Guedens and T.~Jacobson,
Phys. Rev. Lett. \textbf{96}, 121301 (2006)
doi:10.1103/PhysRevLett.96.121301
[arXiv:gr-qc/0602001 [gr-qc]].









\end{thebibliography}
\end{document}